\documentclass[english]{revtex4}
\usepackage[T1]{fontenc}
\usepackage[latin1]{inputenc}
\usepackage{amsmath}
\usepackage{graphicx}
\makeatletter
\usepackage{babel}
\makeatother
\begin{document}
\newcommand{\bc}{\begin{center}}
\newcommand{\ec}{\end{center}}
\newcommand{\bt}{\begin{tabular}}
\newcommand{\et}{\end{tabular}}
\newcommand{\bfr}{\begin{flushright}}
\newcommand{\efr}{\end{flushright}}
\newcommand{\bfl}{\begin{flushleft}}
\newcommand{\efl}{\end{flushleft}}
\newcommand{\cC}{{\cal C}}
\selectlanguage{english}

\title{Optimization and Physics: 
On the satisfiability of random Boolean formulae     }
\author{ Marc M\'ezard}
\affiliation{  LPTMS, Universit\'e de Paris Sud, Orsay, France}

\date{\today}

\begin{abstract}
LECTURE GIVEN AT TH2002:
Given a set of Boolean variables, and some constraints between them,
is it possible to find a configuration of the variables which satisfies
all constraints? This problem, which is at the heart
of combinatorial optimization and computational complexity theory, 
is used as a guide to show the convergence
between these fields and the statistical physics of disordered systems.
New results on satisfiability, both on the theoretical and practical side,
can be obtained thanks to the use of physics concepts and methods.
\end{abstract}
\maketitle

 Combinatorial optimization aims at finding, in a finite set of possible
configurations, the one which minimizes a certain cost function. The famous
example of the traveling salesman problem (TSP) can serve as an illustration:
A  salesman must make a tour through $N$ cities, visiting each city
only once, and coming back at its starting point. The cost function is a
symmetric matrix $c_{ij}$, where $c_{ij}$ is the cost for the travel between
cities $i$ and $j$. A permutation $\pi$ of the $N$ cities gives a tour
$\pi(1)\to \pi(2) \to \pi(3)\to...$. Taking into account the equivalence
between various starting points and the direction of the tour, one sees that
the number of distinct tour is $(N-1)!/2$. For each tour $\pi$, the total
cost is $C=c_{\pi(N)\pi(1)}+\sum_{r=1}^{N-1} c_{\pi(r)\pi(r+1)}$, which can be
computed in $N$ operations. 
The problem is to find the tour $\pi$ with lowest cost.

As can be seen on this example, the basic
ingredients of the optimization problems which will interest us are the
following:
\begin{itemize}
\item An integer $N$ giving the size of the problem (in the TSP, it is
the number of cities).
\item A set of configurations, which typically scales like $\exp(N^\alpha)$.
\item A cost function, which one can compute in polynomial time $O(N^\gamma)$.
\end{itemize}

Let me mention a few examples, beside the TSP \cite{papad}.

In the assignment problem, one is given a set of $N$ persons $i=1,...,N$, a set
of $N$ tasks $a=1,...,N$, a $N \times N$ cost matrix $c$ where $c_{ia}$ is the
cost for having task $ a$ performed by person $i$. An assignment is a
permutation $\pi \in S_N$ assigning task $a=\pi(i)$ to person $i$, and the
problem is to find the lowest cost assignment, i.e. the permutation which
minimizes $C=\sum_{i=1}^N c_{i\pi(i)}$.

In the spin glass problem\cite{MPV}, one is given a set of $N$ spins, which could be
for instance Ising variables $\sigma_i\in\{ \pm 1 \}$, the total energy 
of the configuration is a sum of exchange interaction
energies between all pairs of spins, 
$E(\{\sigma\})=-\sum_{1\le i<j \le N}J_{ij} \sigma_i \sigma_j$,
and one seeks the ground state of the problem, i.e. the configuration 
(among the $2^N$ possible ones) which minimizes the energy.

In physical terms, optimization problems consist in finding ground states.
This task can be non trivial if a system is frustrated, which means that one
cannot get the ground state simply by minimizing the energy locally. This is
typically what happens in a spin glass. In some sense, statistical physics
addresses a more general question. Assuming that the system is at thermal
equilibrium at temperature $T$, every configuration $\cC$ is assigned a
Boltzmann probability, $P(\cC)=\frac{1}{Z} e^{-{ \beta} { E(\cC)}}$. Beside
finding the ground state, one can ask also interesting questions about which
are the typical configurations at a given temperature, like counting them
(which leads to the introduction of entropy), or trying to know if they are
located in one single region of phase space, or if they build up well
separated clusters, as often happens in situation where ergodicity is broken.
The generalization introduced by using a finite temperature, beside leading to
interesting questions, can also be useful for optimization, both from the algorithmic point of
view (for instance this is the essence of the simulated annealing
algorithm\cite{KGV}, which is a general purpose heuristic algorithm that can
be used in many optimization problems), but also from an analytic point of
view \cite{MPV}. Conversely, smart optimization algorithms turn out to be very
useful in the study of frustrated physical systems like spin glasses or random
field models, and the cross-fertilization between these two fields (and also
with the related domain of error correcting codes for information transmission
\cite{codes}) has been very fruitful in the recent years \cite{TCS_issue}.
 
Before proceeding with one such example, let us briefly mention
a few important results in optimization which will
provide the necessary background and motivation.
 One of the great achievements of computer science
is the theory of  computational complexity. It is impossible to present
it in any details here and I will just sketch a few main ideas,
the interested reader can study it for instance in \cite{papad2}.

Within the general framework explained above, we can define three types of
optimization problems: the optimization problem in which one wants to find the
optimal configuration, the evaluation problem in which one wants to compute
the optimal cost (i.e. the ground state energy), the decision problem in which
one wants to know, given a threshold cost $C_0$, if there exists a
configuration of cost less than $C_0$.

One classification of decision problems is based on the scaling with $N$
of the computer  time required to solve them  in the {\it  worst case}.
There are two main complexity classes:
\begin{itemize}
\item { {\bf Class P}\ \ } = polynomial problems: they can be solved in a time
$ t <N^\alpha$. The assignment is an example of a polynomial problem, as is
the spin glass problem in 2 dimensions.
\item{ {\bf Class NP}} = non-deterministic polynomial: Given a 'yes' solution
to a NP problem, it can be checked in polynomial time. Roughly speaking this
means that the energy is computable in polynomial time, so this class contains
a wide variety of problems, including most of the ones of interest in physics.
All problems mentioned above are in NP.
\end{itemize}
One nice aspect of focusing on polynomiality is that it allows to forget about
the details of the definition of $N$, the implementation, language, machine,
etc...: any 'reasonable' such change (for instance one could have used the
number of possible links in the assignment) will change the exponent of $N$
appearing in the computer time of a problem in P, but not transform it into an exponential
behavior. A problem $A$ is said to be at least as hard as a problem $B$ if
there exists a polynomial algorithm which transforms any instance of $B$ into
an instance of $A$.

This allows to define the very important class:
\begin{itemize}
\item { {\bf NP-complete}} A problem   is NP complete if it is at least as hard
as any problem in NP.
\end{itemize}
So the NP-complete are the hardest problems in NP. If one such problem can 
be solved in polynomial time, then all the problems in NP are solved in polynomial
time. Clearly P is contained in NP, but it is not yet known whether
 P = NP , and this is considered as a major challenge.

A great result was obtained in this field by Cook in 1971 \cite{Cook}: 
The satisfiability problem, which we shall describe below,
is NP-complete. Since then hundreds of problems have been shown to belong to
this class, among which the decision TSP or the spin glass in dimension larger
or equal to 3.

The fact that 3-d spin glass is NP-complete while 2-d spin glass is P might
induce one to infer that NP-completeness is equivalent with the existence of a
glass transition. This reasoning is too naive and wrong; the reason is that
the complexity classification relies on a worth-case analysis, while
physicists study the behavior of a typical sample. The development of a
typical case complexity theory has become a major goal \cite{average}, also motivated by the
experimental observations that many instances of NP-complete problems are easy
to solve\cite{TCS_issue}.

One way of addressing this issue of a typical case complexity is to define
a probability measure on the instances (= the 'samples')
of the optimization problem which one is considering.
Typical examples are:
\begin{itemize}
\item TSP with independent random points uniformly distributed in the unit square
\item assignment with independent affinities uniformly distributed on $[0,1]$
\item CuMn spin glass  at one percent Mn
\end{itemize}
Once this measure has been defined, one is interested in the properties of the
{ generic} sample in the $N \to \infty$ limit. In most cases, global
properties (e.g. extensive thermodynamic quantities, among which the ground
state energy), turn out to be self averaging. This means for instance that the
distribution of the ground state energy density becomes more and more peaked
around an asymptotic value in the large $N$ limit: almost all samples have the
same ground state energy density. In this situation, a statistical physics
analysis is appropriate. Early examples of the use of statistical physics in
such a context are the derivation of bounds for the optimal length of a
TSP\cite{MVan},  the exact prediction of the ground state energy in the
random assignment problem defined above, where the result
$E_0=\frac{\pi^2}{6}$, derived in 1985 through a replica
analysis\cite{MP_match},   was recently confirmed rigorously by Aldous
\cite{Aldous}, or the link between spin glasses and graph partitioning
\cite{FuAnd}.

As statistical physics can be quite powerful at understanding the generic
structure of an optimization problem, one may also hope that it can help
finding better optimization algorithms. A successful example which was
developed recently is the satisfiability problem, to which we now turn.

As we have seen, satisfiability is a core problem in optimization and
 complexity theory. It is defined as follows \cite{Hayes}: A configuration is
 a set of $N$ Boolean variables $x_i \in \{0,1\} \quad i=1,\ldots,N$. One is
 given $M$ constraints which are clauses, meaning that they are in the form of
 an OR function of some variables or their negations. For instance:  $x_1 \vee
 x_{27} \vee \bar x_3$, $ \bar x_{11} \vee x_2$, are clauses (notation: $\bar
 x_3$ is the negation of $x_3$). So the clause $x_1 \vee x_{27} \vee \bar x_3$
 is satisfied if either $x_1=1$, or $x_{27}=1$, or $x_3=0$ (these events do
 not exclude each other). The satisfiability problem is a decision problem. It
 asks whether there exists a choice of the Boolean variables such that all
 constraints are satisfied (we will call it a SAT configuration). This is a
 very generic problem, because one can see it as finding a configuration of
 the $N$ Boolean variables which satisfies the logical proposition built from
 the AND of all the clauses (in our example{ $(x_1 \vee x_{27} \vee \bar x_3)
 \wedge (\bar x_{11} \vee x_2) \wedge \ldots$}), and any logical proposition
 can be written in such a 'conjunctive normal form' .

Satisfiability is known to be NP complete if it contains clauses of length
 $\ge 3$, but common sense and experience show that the problem can often be
 easy; for instance if the number of constraints per variable
 $\alpha=\frac{M}{N}$ is small, the problem is often SAT, if it is large, the
 problem is often UNSAT.

This behavior can be characterized quantitatively by looking at the typical
complexity of the random 3-SAT problem, defined as follows. Each clause
involves exactly three variables, chosen randomly in $\{ x_1,..,x_N\}$; a
variable appearing in the clause is negated randomly with probability $1/2$.
This defines the probability measure on instances for the random 3-SAT
problem. The control parameter is the ratio Constraints/Variables $
\alpha=\frac{M}{N}$ .

The properties of random 3-SAT have been first investigated numerically
\cite{KirkSel,Crawford}, and exhibit a very interesting threshold
phenomenon at $\alpha_c \sim 4.26$: a randomly chosen sample is generically
SAT for $\alpha<\alpha_c$ (meaning that it is SAT with probability $1$ when $N
\to \infty$), generically UNSAT for $\alpha>\alpha_c$. The time used by the
best available algorithms (which have an exponential complexity) to study the
problem also displays an interesting behavior: For $\alpha$ well below
$\alpha_c$, it is easy to find a SAT configuration; for $\alpha$ well above
$\alpha_c$, it is relatively easy to find a contradiction in the constraints,
showing that the problem is UNSAT. The really difficult region is the
intermediate one where $\alpha \sim \alpha_c$, where the computer time
requested to solve the problem is much larger and increases very fast with
system size. A lot of important work has been done on this problem, to
establish the existence of a threshold phenomenon, give upper and lower bounds
on $\alpha_c$, show the existence of finite size effects around $\alpha \sim
\alpha_c$ with scaling exponents. We refer the reader to the literature
\cite{KirkSel,Crawford,MZKST,bounds_1,bounds_2,bounds_3,bounds_4}, and
just here extract a few crucial aspects for our discussion. The threshold
phenomenon is a phase transition, and the neighborhood of the transition is
the region where the algorithmic problem becomes really hard.

This relationship between phase transition and complexity makes a
statistical physics analysis of this problem particularly interesting. 
Monasson and Zecchina have been the first ones to recognize this importance
and to use statistical physics methods for an analytic
study of the random 3-SAT problem \cite{MoZe_prl,MoZe_pre}.
Basically this problem  falls into the broad category of 
spin glass physics. This is immediately seen through the following
formulation. To make things look more familiar, physicists like 
to introduce for each Boolean variable $x_i\in \{0,1\}$
an Ising spin $\sigma_i \in\{-1,1\}$ through $x_i= \frac{1+\sigma_i}{2}$.
A satisfiability problem like 
\begin{equation}
{ (x_1 \vee x_{27} \vee \bar x_3) }{\wedge }{  (\bar x_{11} \vee x_3 \vee x_2)
 }{\wedge}{  \ldots }
\end{equation}
 can be written in terms of an energy function, where the energy is equal to one
for each violated clause. Explicitly, in the above example, one would have
\begin{equation}
 E= \frac{1+\sigma_1}{2}\; \frac{1+\sigma_{27}}{2}\; \frac{1-\sigma_3}{2}
\;{ +} \;\frac{1-\sigma_{11}}{2}\; \frac{1+\sigma_3}{2}\; \frac{1+\sigma_2}{2} \;{ +} \; \ldots
\end{equation}
This is clearly a problem of interacting spins, with 1,2, and 3 spin interactions,
disorder (in the choice of which variable interacts with which),
and frustration (because some constraints are antagonist).
More technically, the problem has a special type of three-spin interactions on a random hyper-graph.

Using the replica method, Monasson and Zecchina first showed the existence
of  a phase transition within the
replica symmetric approximation, at $\alpha_c=5.18$, then showed that
replica symmetry must be broken in this problem.
 Some  variational approximation to describe the replica symmetry breaking
effects were developed in particular in \cite{BMW,FLRZ}. 

Recently, in a collaboration with G.Parisi and R. Zecchina
\cite{MEPAZE,MZ_pre}, we have developed a new approach to the statistical
physics of the random 3-SAT problem using the cavity method. While the cavity
method had been originally invented to deal with the SK model where the
 interactions are of infinite range \cite{MPV}, it was later adapted to
problems with 'finite connectivity', in which a given variable interacts with
a finite set of other variables. While this is easily done for systems which
are replica symmetric (like the assignment, or the random TSP with independent
links), it turned out to be considerably more difficult to develop the
corresponding formalism and turn it into a practical method, in the case where
replica symmetry is broken. This has been done in the last two years in joint
works with G.Parisi \cite{MP_Bethe}, and has opened the road to the study of
finite connectivity optimization problems with replica symmetry breaking like
random K-sat. Curiously, while the cavity method is in principle equivalent to
the replica method (although it proceeds through direct probabilistic analysis
instead of using an analytic continuation in the number of replicas),
it turns out that it is easier to solve this problem with the cavity method.

\begin{figure}
\includegraphics[width=9.cm]{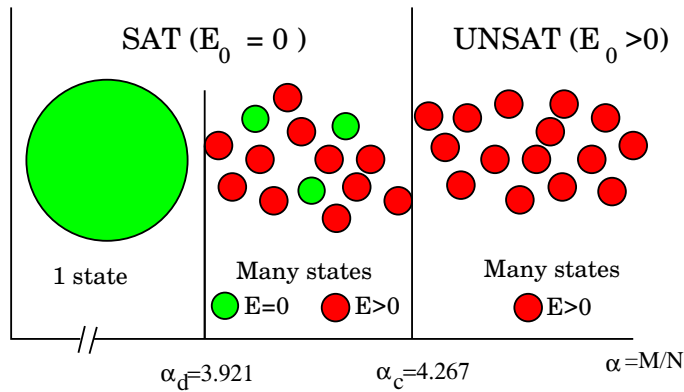}
\caption{A pictorial description of the phase diagram for random 3-SAT
obtained in the cavity method: clusters of SAT (in green/light gray) or UNSAT (in red/dark grey)
configurations. One finds three qualitatively different phases, the
 EASY-SAT phase for $\alpha<\alpha_d$, the HARD-SAT phase for
$\alpha_d<\alpha<\alpha_c$, the UNSAT phase obtained for $\alpha_c<\alpha$
(see text)
}
\label{phdi}
\end{figure}

The  analytic 
study of \cite{MEPAZE,MZ_pre} for the random 3-SAT problem
at zero temperature  shows the
following phase diagram (see fig.\ref{phdi}):
\begin{itemize}
\item
For $\alpha<\alpha_d=3.921$, the problem is generically SAT; the
solution can be found relatively easily, because the space
of SAT configurations builds up a single big connected cluster.
A $T=0$ Metropolis algorithm, in which one proposes to flip a randomly chosen variable, and accepts the change iff the number of violated constraints in the new
configuration is less or equal to the old one, is able to find a SAT configuration.
We call this the EASY-SAT phase
\item
For $\alpha_d<\alpha<\alpha_c=4.267$, the problem is still generically SAT,
but now it becomes very difficult to find a solution (we call this
the HARD-SAT phase). The reason is that
the configurations of variables which satisfy all constraints build up
some clusters. Each such cluster, which we call  a 'state', is well separated from
the other clusters (passing from one to the other requires flipping an extensive
number of variables). But there also exist many ``metastable states'': starting
from a random configuration, a local descent algorithm will get trapped
in some cluster of configurations with a given finite energy
 (they all have the same number
of violated clauses, and it is impossible to get out of this cluster
towards lower energy configurations through any descending 
sequence of one spin flip moves). The number of SAT clusters ${\cal N}$ 
 is exponentially 
large in $N$, it behaves as ${\cal N}\sim \exp(N \Sigma)$;
 but the number of metastable clusters is also exponentially large 
with a larger growth rate, behaving like $ \exp(N \Sigma_{ms})$
with $\Sigma_{ms}> \Sigma$. The most numerous metastable clusters,
which have an energy density $e_{th}$, will trap all local descent algorithms
(zero temperature Metropolis of course, but also simulated annealing,
unless it is run for an exponential time).

\item For $\alpha>\alpha_c$, the problem is typically UNSAT.
The ground state energy density $e_0$ is positive. Finding 
a configuration with lowest energy 
 is also very difficult because of the proliferation of metastable states.
\end{itemize}

\begin{figure}
\includegraphics[width=10.cm]{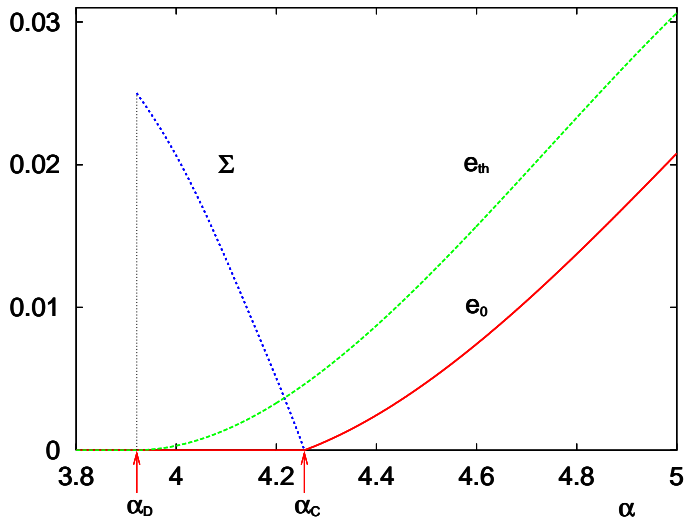}
\caption{ Thermodynamic quantities for the random 3-SAT problem:
$e_0$ is the ground state energy density (minimal number of violated clauses per variable);
$e_{th}$  is the energy density  of the most numerous metastable states,which trap the local descent algorithms;
$\Sigma$ is the complexity of SAT states with $E=0$ }
\label{phdi2}
\end{figure}

A more quantitative description of the thermodynamic quantities in the various
phases is shown in fig.\ref{phdi2}. The most striking result is the existence
of an intermediate SAT phase where clustering occurs. A hint of such a
behavior had been first found in a sophisticated variational one step replica
symmetry breaking approximation developed in \cite{BMW}; however this
approximation predicted a second order phase transition (clusters separating
continuously), while we now think that the transition is discontinuous: an
exponentially large number of macroscopically separated clusters appears
discontinuously at $\alpha=\alpha_d$. Another point which should be noticed is
the fact that the complexity, and the energy $e_{th}$ in the HARD-SAT phase
are rather small: around $e_{th} \sim 3 \ 10^{-3}$ violated clause per variable
for $\alpha=4.2$. This shows that until one reaches problems with at least a
few thousands variables, one cannot feel the true complexity of the problem.
This can explain why the existence of the intermediate phase went unnoticed in
previous simulations.

\begin{figure}
\includegraphics[width=7.cm]{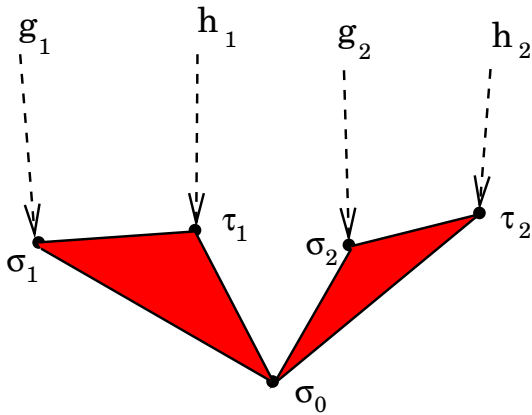}
\caption{The basic iterative structure of survey propagation. This subgraph is
a part of a 3-SAT problem, in which the variable $\sigma_0$ belongs
to three 3-clauses, involving respectively the variables $\{\sigma_1,\tau_1\}
,\{\sigma_2,\tau_2\},\{\sigma_3,\tau_3\}$. When the clause 3 is switched off,
the local cavity field survey $P_0^e(h)$ on spin $\sigma_0$ can be computed
in terms of the cavity field surveys on each of the spins $\sigma_1,\tau_1
,\sigma_2,\tau_2$.}
\label{iterfig}
\end{figure}

The second type of results found in \cite{MEPAZE,MZ_pre,BMZ} is a
  new class of algorithms dealing with the many states structure.
Basically this algorithm amounts to using the cavity equations on one given sample.
Originally the cavity method was developed to handle a statistical 
distribution of samples. For instance in the random  3-SAT case its
basic strategy is to add one extra variable and connect it randomly to
a number of new clauses which has the correct statistical distribution.
In the large $N$ limit, the statistics of the local field on the
new variable should be identical to the statistics of the local fields on 
any other variable in the absence of the new one. It turns out that
this strategy can be adapted to study a single sample: one considers
a given clause $a$, which involves three variables $\sigma_1,\sigma_2,\sigma_3$.
The cavity field on $\sigma_1$ is the field felt by $\sigma_1$
in the absence of $a$. In the case where there exist many states, there is one
such field for each possible state of the system, and the order parameter is the survey of these fields, in all the states of fixed energy density $e$:

\begin{equation}
 P_0^e(h)= C^t \sum_\alpha \delta\left(h_0^\alpha-h\right) \delta
\left(\frac{E^\alpha}{N}-e\right)
\end{equation}
One can then write a recursion recursion between these surveys.
Looking for instance at the structure of  fig.\ref{iterfig}, one gets the following 
iteration equation:

\begin{eqnarray}\nonumber
 P_0^e(h)&=& C^t \int P_{\sigma_1}^e(g_1)dg_1 \;  P_{\tau_1}^e(h_1)dh_1 
\;P_{\sigma_2}^e(g_2)dg_2 \;  P_{\tau_2}^e(h_2)dh_2 \\
&&
\delta(h-f(g_1,h_1,g_2,h_2))
\exp\left(-\frac{d \Sigma}{d e} w(g_1,h_1,g_2,h_2)\right)
\label{speq}
\end{eqnarray}

The function $f$ just computes the value of the new cavity field
on $\sigma_0$ in terms of the four cavity fields $g_1,h_1,g_2,h_2$.
It is easily computed by considering the statistical mechanics
problem of the five-spin system $\{ \sigma_0,\sigma_1,\tau_1
,\sigma_2,\tau_2\}$ and summing over the 16 possible states of the 
spins $\sigma_1,\tau_1,\sigma_2,\tau_2$. The function $w$
computes the free energy shift induced by the addition of $\sigma_0$ 
to the system with the four spins  $\sigma_1,\tau_1,\sigma_2,\tau_2$.
The exponential reweighting term in (\ref{speq}) is the crucial piece of 
survey propagation: it appears because one considers the survey
at a fixed energy density $e$. As the number of states
at energy $E=Ne+\delta E$ increases in $\exp(y \delta E)$,
 where  $y=\frac{d \Sigma}{d e}$, this favors the states with a large negative
energy shift.

The usual cavity method for the random 3-SAT problem determines the
probability, when a new variable is added at random, that its survey 
$P_0^e(h)$ is a given function $P(h)$: the order parameter is thus
 a functional, the probability of a probability.
 Because the fields are distributed on integers, this object is not as
 terrible as it may look and
it is  possible to solve the equation and compute 
the 'complexity curve' $\Sigma(E)$, giving the phase diagram described above.

The algorithm for one given sample basically iterates the survey
propagation equation on one given graph. It is a message passing 
algorithm which  can be seen as a generalization
of the belief propagation algorithm familiar to computer scientists\cite{pearl}.
The  belief propagation is a propagation of local magnetic fields,
which is equivalent to using a Bethe approximation \cite{yed}.
Unfortunately, it does not converge in the hard-SAT region because 
various subparts of the graph tend to settle in distinct states,
and there is no way to globally choose a state. In this region,
the survey propagation, which propagates the information on 
the whole set of states (in the form of a histogram), does 
converge.

\begin{figure}
\includegraphics[width=7.cm]{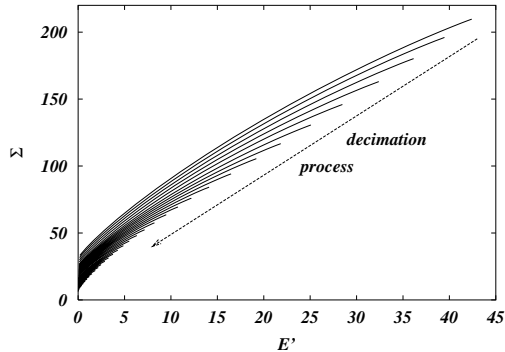}
\caption{The total complexity (= the $\ln$ of the number of states)
as a function of the total energy (= the number of violated clauses)
for one given sample with  $N=10^4$ variables and $M=4.2 \; 10^4$ 
constraints. The top curve is the original complexity.
The next curves are the various complexities obtained 
for the decimated samples, plotted here every $200$ decimations.
One sees a global decrease of the number of metastable
states, and also a global decrease
of the threshold energy. In the end the problem has no more metastable states and 
can be solved by simple algorithms.
}
\label{fig_decim}
\end{figure}

Based on the surveys, one can detect some strongly biased spins,
which are fixed to one in almost all SAT configurations.
The  ``Survey Inspired Decimation''  (SID) 
 algorithm fixes the spin which is most biased,
then it  re-runs the survey propagation on the smaller sample so obtained, 
and then iterates... An example of the evolution of
the complexity as a function of the decimation is shown in \ref{fig_decim}.
This algorithm has been tested in the hard SAT phase. It easily
solves the 'large'  benchmarks of random 3sat at 
$\alpha=4.2$ with $N=1000,2000$ available at \cite{SATLIB}.
It turns out to be able to solve
 typical random 3-SAT problems with  up to $N=10^7$ at  $\alpha=4.2$
in a few hours on a PC, which makes it much better than available algorithms.
The main point is that the set of surveys contains a lot of
detailed information on a given sample and can probably be used to find
many new algorithms, of which SID is one example.

To summarize, the recent statistical physics approaches to
the random 3-SAT problem give the following results:
\begin{itemize}
\item
An analytic result for the phase diagram of the generic samples
\item
An explanation for the slowdown of algorithms near  $\alpha_c=4.267$:
this is  due to the existence of a
{ HARD-SAT phase} at $\alpha \in [3.921,4.267]$, with exponentially many
metastable states.
\item
An algorithm for single sample analysis\cite{web}:  Survey propagation  converges
and yields
very non trivial information on the sample. It can be used
for instance to decimate the problem and get an efficient SAT-solver in 
the hard-SAT region.
\end{itemize}

This whole set of results calls for a lot of developments in many directions.

On the analytical side, the cavity method results quoted above are believed to
be correct, but they are not proven rigorously. It would be very interesting
to turn these computations into a rigorous proof.
A very interesting step in this direction was taken recently
by Franz and Leone who showed that the result for the
critical threshold $\alpha_c$ obtained by the cavity method
on random K-SAT with $K$ even actually give a rigorous upper bound to
the critical $\alpha_c$ \cite{FraLeo}.
The whole
 construction of the cavity method with the clustering structure,
the many states and the resulting reweighting,  has actually
been  checked versus rigorous computations on a variant of the SAT problem, the random XORSAT problem, where rigorous computations \cite{xorsat}
have confirmed the validity of the approach.

On the numerical side, one needs to develop convergence proofs for survey
propagation, and to derive the generalization of the algorithm for the case in
which there exists local structures in the interaction graphs. This will
amount to generalizing from a Bethe like approximation (with many states) to a
cluster variational method with larger clusters (and with many states).

The techniques can also be extended to other optimization problems like 
coloring \cite{colouring}.
Beside the problems mentioned here, there exist many other fascinating
problems of joint interests to physicists and computer scientists, like
e.g the dynamical behavior of algorithms in optimization
or error correcting codes, which I cannot survey here. Let me just quote a few recent
references to help the readers through the corresponding bibliography
\cite{cocmon,global_algorithm,FLMR,MonZec}.

{Acknowledgements}
It is a pleasure to thank all my collaborators over the years on
various topics at the interface between statistical physics and
optimization. I am particularly thankful to G. Parisi and
R. Zecchina for the wonderful collaborations over the last three years which
lead to the works on the cavity method and the satisfiability problem.


\begin{thebibliography}{1}
\bibitem{papad} C.H. Papadimitriou and K. Steiglitz, 
{\it Combinatorial Optimization}, Dover (Mineola, N.Y., 1998).

\bibitem{MPV} M\'ezard, M., Parisi, G., $\&$ Virasoro, M.A. Spin Glass
Theory and Beyond, World Scientific, Singapore (1987).

\bibitem{KGV} Kirkpatrick, S., {Gelatt Jr.}, C.~D. $\&$ Vecchi,
M.~P.  Optimization by simulated annealing, {\it Science}{\bf} 220
671--680(1983).  {\v C}erny, V., Thermodynamical approach to the
traveling salesman problem: An efficient simulation algorithm, {\it
J.~Optimization Theory Appl.}{\bf 45} 41 (1985).


\bibitem{codes} See e.g. A. Montanari, {\it The glassy phase of Gallager codes}
cond-mat/0104079, and references therein.

\bibitem{TCS_issue} Dubois O. , Monasson R., Selman B. $\&$ Zecchina
R.  (Eds.), Phase Transitions in Combinatorial Problems,
Theoret. Comp. Sci. {\bf 265} (2001).



\bibitem{papad2} C.H. Papadimitriou, {\it Computational complexity}
(Addison-Wesley, 1994)

\bibitem{Cook} Cook, S.  The complexity of theorem--proving procedures,
in: Proc. 3rd Ann. ACM Symp. on Theory of Computing,
Assoc. Comput. Mach., New York, 1971, p. 151.

\bibitem{average}  Levin L.A., Average case complete problems, SIAM
J. Comput., {\bf 14} (1): 285-286 (1986); Ben-David S., Chor B., Goldreich O. $\&$ Luby M.,
On the theory of average case complexity, JCSS {\bf 44}, 193-219;
 Gurevich Y., Average Case completeness, JCSS {\bf
42}, 246-398 (1991).

\bibitem{MVan}, M. M\'ezard and J. Vannimenus, {\it On the statistical
mechanics of optimization problems of the traveling salesman type},
J. Physique Lett.  {\bf 45} (1984) L1145.

\bibitem{MP_match} M. M\'ezard and G. Parisi, Replicas and Optimization, 
J.Phys.Lett. {\bf 46} (1985) L771.

\bibitem{Aldous} D. Aldous, The zeta(2) Limit in the Random Assignment
Problem, Random Structures and Algorithms {\bf 18} (2001) 381-418.

\bibitem{FuAnd} Fu Y. and Anderson P. W., Application of Statistical
Mechanics to NP-Complete Problems in Combinatorial Optimization,
J. Phys. A 19, 1605--1620 (1986).

\bibitem{Hayes} A nice  pedagogical introduction is B. Hayes
{\it I can't get no satisfacion}, American Scientist 
{\bf 85}  (1997)  108.


\bibitem{KirkSel} S. Kirkpatrick, B. Selman, Critical Behaviour in the
satisfiability of random Boolean expressions, Science 264, 1297 (1994)

\bibitem{Crawford} Crawford J.A. $\&$ Auton L.D., Experimental results
on the cross-over point in random 3-SAT, Artif. Intell. {\bf 81},
31-57 (1996).
\bibitem{MZKST} R. Monasson, R. Zecchina, S. Kirkpatrick, B. Selman
and L.  Troyansky, Nature (London) {\bf 400}, 133 (1999).

\bibitem{bounds_1} A. Kaporis, L. Kirousis, E. Lalas, The
probabilistic analysis of a greedy satisfiability algorithm, in {\it
Proceedings of the 4th European Symposium on Algorithms} (ESA 2002),
to appear in series: Lecture Notes in Computer Science, Springer;

\bibitem{bounds_2} D. Achlioptas, G. Sorkin, {\it 41st Annu. Symp. of
Foundations of Computer Science, IEEE Computer Soc. Press}, 590 (Los
Alamitos, CA, 2000).

\bibitem{bounds_3} J. Franco, Results related to threshold phenomena
research in satisfiability: lower bounds, Theoretical Computer Science
{\bf 265}, 147 (2001)

\bibitem{bounds_4} O. Dubois, Y. Boufkhad, J. Mandler, Typical random
3-SAT formulae and the satisfiability threshold, in {\it Proc. 11th
ACM-SIAM Symp. on Discrete Algorithms}, 124 (San Francisco, CA, 2000).

\bibitem{MoZe_prl} Monasson, R.  $\&$ Zecchina, R. Entropy of the
{K}-satisfiability problem, {\it Phys. Rev.  Lett.} {\bf 76}
3881--3885(1996).

\bibitem{MoZe_pre} Monasson, R.  $\&$ Zecchina, R., Statistical
mechanics of the random {K-S}at problem, {\it Phys. Rev. } {\bf E 56}
1357--1361 (1997).

\bibitem{BMW} Biroli, G., Monasson, R.  $\&$ Weigt, M. A Variational
description of the ground state structure in random satisfiability
problems, {\it Euro. Phys. J. } {\bf B 14} 551 (2000).

\bibitem{FLRZ} S. Franz, M. Leone, F. Ricci-Tersenghi, R.  Zecchina,
Exact Solutions for Diluted Spin Glasses and Optimization Problems,
{\it Phys. Rev. Lett.} {\bf 87}, 127209 (2001).

\bibitem{MEPAZE} M. M{\'e}zard, G. Parisi and R. Zecchina, Science
297, 812 (2002) (Sciencexpress published on-line 27-June-2002;
10.1126/science.1073287)

\bibitem{MZ_pre} {\it Random 3-SAT: from an analytic solution to a new
efficient algorithm}, M. M{\'e}zard and R. Zecchina, Phys.Rev. E {\bf 66}
(2002) 056126.

\bibitem{BMZ} {\it Survey propagation: an algorithm for satisfiability},
 A. Braunstein, M. M\'ezard, R. Zecchina, http://fr.arXiv.org/abs/cs.CC/0212002.

\bibitem{MP_Bethe}
M{\'e}zard, M. $\&$ Parisi, G. The Bethe lattice
spin glass revisited.  {\it Eur.Phys. J. B} {\bf 20}, 217--233 (2001).
 M{\'e}zard, M. $\&$ Parisi, G. The cavity method at
zero temperature, J. Stat. Phys.  in press.

\bibitem{pearl} J. Pearl, {\it Probabilistic Reasoning in Intelligent Systems},
2nd ed. (San Francisco, MorganKaufmann,1988).

\bibitem{yed} J.S. Yedidia, W.T. Freeman and Y. Weiss,
Generalized Belief Propagation, in {\it Advances in Neural Information
Processing Systems 13} eds. T.K. Leen, T.G. Dietterich, and V. Tresp,
MIT Press 2001, pp. 689-695.

\bibitem{SATLIB} Satisfiability Library: www.satlib.org/


\bibitem{FraLeo} S. Franz and M. Leone, Replica bounds for optimization problems
and diluted spin systems, preprint cond-mat/0208280, available at http://fr.arXiv.org

\bibitem{xorsat} M. M\'ezard, F. Ricci-Tersenghi, R. Zecchina, {\it
Alternative solutions to diluted p-spin models and XOR-SAT problems},
cond-mat/0207140, (2002). S. Cocco, O. Dubois, J. Mandler, R. Monasson, {\it
Rigorous decimation-based construction of ground pure states for spin
glass models on random lattices}, cond-mat/0206239 (2002).
\bibitem{colouring}{\it Coloring random graphs}, R. Mulet, A. Pagnani, M. Weigt, R. Zecchina, cond-mat/0208460.
\bibitem{cocmon}{\it Restart method and exponential acceleration of random 3-SAT instances resolutions: a large deviation analysis of the
     Davis-Putnam-Loveland-Logemann algorithm},
S. Cocco and R. Monasson, cond-mat/0206242
\bibitem{global_algorithm} {\it Complexity transitions in global
algorithms for sparse linear systems over finite fields}, J. Phys. A
35, 7559 (2002), A. Braunstein, M. Leone, F. Ricci-Tersenghi,
R. Zecchina.

\bibitem{FLMR} {\it The Dynamic Phase Transition for Decoding Algorithms},
S. Franz, M. Leone, A. Montanari, F. Ricci-Tersenghi,
cond-mat/0205051.

\bibitem{MonZec} {\it Boosting search by rare events}, A. Montanari, R. Zecchina,
cond-mat/0112142.




\bibitem{web} {\tt www.ictp.trieste.it/$\tilde{~}$zecchina/SP}



\end{thebibliography}
\end{document}